\documentstyle[12pt]{article}                    
\newcommand{\beq}{\begin{equation}}             
\newcommand{\eeq}{\end{equation}}               
\newcommand{\bqry}{\begin{eqnarray}}            
\newcommand{\eqry}{\end{eqnarray}}              
\newcommand{\bqryn}{\begin{eqnarray*}}          
\newcommand{\eqryn}{\end{eqnarray*}}            
\newcommand{\preprint}[1]{\begin{table}[t]      
            \begin{flushright}                  
            \begin{large}{#1}\end{large}        
            \end{flushright}                    
            \end{table}}                        
\newcommand{\PD}[2]                             
    {\frac{\partial^{#2}}{\partial #1^{#2}}}    
\begin{document} 
\preprint{LA-UR-00-428} 
\title{On the Bragg, Leibfried, and \\ Modified Leibfried Numbers} 
\author{\\ Leonid Burakovsky\thanks{E-mail: BURAKOV@T5.LANL.GOV}, \
Dean L. Preston\thanks{E-mail: DEAN@LANL.GOV}, \
and Richard R. Silbar\thanks{E-mail: SILBAR@WHISTLESOFT.COM. Also at 
WhistleSoft, Inc., 
Los Alamos, NM 87544, USA}
 \\  \\ 
Los Alamos National Laboratory \\ Los Alamos, NM 87545, USA }
\date{ }
\maketitle
\begin{abstract} 
The Bragg, Leibfried, and modified Leibfried numbers are defined in the 
context of a theory of dislocation-mediated melting, and their values are 
determined from the properties of the dislocation ensemble at the melting 
temperature. The approximate numerical coincidence of the Bragg and modified  
Leibfried numbers is explained. The parameter $K$ in the definition of 
the modified Leibfried number is shown to be the natural logarithm of the 
effective coordination number. Our analysis reveals that the Bragg number 
can be considered an elemental constant, in contrast to the Leibfried and 
modified Leibfried numbers. 
\end{abstract}
\bigskip 
\centerline{{\it Key words:} melting, dislocation, Bragg, Leibfried, modified 
Leibfried} 

\hspace*{1.1cm} PACS: 61.72.Bb, 61.72.Lk, 64.70.Dv, 64.90.+b 
\bigskip


In Leibfried's study of melting \cite{Leib} he observed that for a number of 
metals (specifically Al, Ag, Au, Cu, Pd and Pb) 
\beq
L\equiv RT_m/GV\simeq 0.042,
\eeq
where $R$ is the gas constant, $T_m$ is the melting temperature, $G$ is the 
ambient shear modulus, and $V$ is the molar volume. The factor $RT_m$ is an 
approximation to the heat of fusion, $L_m,$ which follows from Richard's rule 
\cite{Gsch} 
\beq
L_m/T_m=\triangle S_m\simeq 2.0\;{\rm e.u.}\approx R 
\eeq
(1 e.u.\ $=1$ cal/mol $^\circ $K), since the numerical value of $R$ is 1.987 
e.u. Bragg \cite{Bragg}, on the other hand, noted that 
\beq 
B\equiv L_m/GV\simeq 0.034 
\eeq 
for a few metals (specifically Al, Ag, Au, Co, Cu, Fe, Ni and Pb). 

Although $L,$ the Leibfried number, and $B,$ the Bragg number, seem to be 
intrinsically related, through Richard's rule, there is a relatively large 
discrepancy between their values. The importance of both numbers in the 
calculation of the size factors caused Gschneidner to reanalyze them in his 
review \cite{Gsch}. First, he noticed that Richard's rule may not be a valid 
approximation. In fact, Stull and Sinke \cite{SS} had earlier used two 
different values for estimating heats of fusion, 2.3 e.u.\ for elements which 
crystallize as face-centered cubic (fcc) or hexagonal close-packed (hcp) 
structures, and 1.9 e.u.\ for elements which have body-centered cubic (bcc) 
structure. Second, Gschneidner introduced the modified Leibfried number, 
$L',$ which differs from the Leibfried number $L$ in that the term $RT_m$ is 
replaced by the term $KT_m.$ The value of $K$ depends on the crystal structure 
of the element just below its melting point: 2.29 e.u.\ for fcc or hcp metals, 
1.76 e.u.\ for bcc metals, etc. Finally, he concluded that the {\it modified} 
Leibfried and Bragg numbers do agree within uncertainties. 

In this Technical Note we offer an explanation for the numerical values of 
both $L'$ and $B$ in a theory of dislocation-mediated melting. We obtain the 
numerical values of $B,$ $L$ and $L'$ from the properties of the dislocation 
ensemble at the melting point. We also explain the approximate numerical 
coincidence of the values of $L'$ and $B.$  

In our previous study of melting as a dislocation-mediated phase transition 
on a lattice \cite{prev1,prev2}, we obtained two relations: 
\beq 
k_BT_m=\frac{\lambda \kappa Gv_{WS}}{8\pi \ln z'}\ln \left( 
\frac{\alpha ^2}{4b^2\rho (T_m)}\right) ,
\eeq 
\beq 
L_m=\frac{1}{\lambda }b^2\rho (T_m)RT_m\ln z',
\eeq 
where $\rho (T_m)$ is the critical dislocation density at melt, $v_{WS}$ is 
the Wigner-Seitz volume, $1/\kappa =(1-\nu /2)\pm \nu /2\approx 5/6,$ $\nu $ 
being the Poisson ratio, $b$ is the shortest perfect-dislocation Burgers 
vector, $\lambda \equiv b^3/v_{WS}\approx 4/3,$ and $\alpha =2.9$ accounts 
for nonlinear effects in a dislocation core. Also, $z',$ the effective 
coordination number for a dislocation as a random walk on the lattice, 
satisfies $z'\leq z,$ where $z$ is the coordination number of the lattice. 
These relations hold to $\sim 20$\% accuracy for those elements of the 
Periodic Table (more than half) for which sufficient data exist. 

It then follows that the Bragg number is 
\beq
B=\frac{\kappa }{8\pi }\;\!b^2\rho (T_m)\ln \left( \frac{\alpha ^2}{
4b^2\rho (T_m)}\right) .
\eeq 
We have found \cite{prev2} that the critical dislocation density at melt 
is an elemental constant with an approximate numerical value of $2/3b^2$ for 
three-quarters of the Periodic Table. It corresponds to the situation where, 
on average, half of the atoms are within a dislocation core at melt. Hence, 
the Bragg number must be an elemental constant, and its approximate numerical 
value is, in view of Eq.\ (6) with the numerical values of the parameters 
quoted above, 
\beq 
B\approx \frac{0.92}{8\pi }\approx 0.037, 
\eeq 
in agreement with Bragg's original estimate. We discuss this point in more 
detail below. 

Similarly, the Leibfried number is 
\beq 
L=\frac{\lambda \kappa }{8\pi \ln z'}\ln \left( \frac{\alpha ^2}{
4b^2\rho (T_m)}\right) .
\eeq 
It depends, through $\ln z',$ on the crystal structure of the solid phase 
of the element from which melting occurs, in agreement with Gschneidner's 
observation. It is, however, possible to define modified Leibfried numbers 
that approximately coincide numerically with the Bragg number \cite{Gsch}: 
\beq 
L'\equiv \frac{K}{R}\;\!L,
\eeq 
where $K$ may be defined in either of two ways: 
\beq 
K=\frac{2b^2\rho (T_m)\ln z'}{\lambda }\;{\rm e.u.} 
\eeq 
or 
\beq 
K=\ln z'\;{\rm e.u.} 
\eeq 
In the first case, 
\beq
L'=\frac{\kappa }{4\pi }\;\!b^2\rho (T_m)\ln \left( \frac{\alpha ^2}{
4b^2\rho (T_m)}\right) \frac{{\rm e.u.}}{R} 
\eeq
will approximately coincide numerically with $B$ given in Eq.\ (6), since 
the numerical value of $R$ in units of e.u.\ is very close to 2. In the 
second case, 
\beq 
L'=\frac{\lambda \kappa }{8\pi }\ln \left( \frac{\alpha ^2}{4b^2\rho (T_m)}
\right) \frac{{\rm e.u.}}{R} 
\eeq  
will again approximately coincide numerically with $B$ given in (6), since 
$R\approx 2$ e.u., $b^2\rho (T_m)\approx 2/3$ and $\lambda \approx 4/3.$ 


In the first case, the numerical values of $K$ in e.u., according to Eq.\ 
(10), are 2.26 for fcc and hcp metals (for which $\lambda =\sqrt{2}$ and 
$z=12)$ and 1.99 for bcc metals (for which $\lambda \approx 1.3$ and $z=8),$ 
where we have taken $b^2\rho (T_m)=2/3$ and $z'=z-1$ \cite{prev1,prev2}. These 
values of $K$ are consistent with those presented by Gschneidner, 2.29 and 
1.76, respectively. In the second case, the corresponding values for $z'=z-1$ 
are 2.40 for fcc and hcp metals and 1.95 for bcc metals, and for $z'=z-2$ 
these values are 2.30 for fcc and hcp metals and 1.79 for bcc metals. It is 
therefore seen that in the second case the choice $z'=z-2$ leads to better 
agreement between the values of $K$ calculated from Eq.\ (11) and those given 
by Gschneidner. This is not sufficient, however, to conclude that the choice 
$z'=z-2$ is better than the choice $z'=z-1$ used in \cite{prev1,prev2}, since 
Gschneidner \cite{Gsch} did not specify the criteria for his choice of the 
values of $K.$ Equation (5) with $z'=z-1,$ for example, explains the 
Stull-Sinke rule, since it gives $L_m/T_m=2.25$ e.u.\ for fcc and hcp metals, 
and 1.98 e.u.\ for bcc metals, in good agreement with the values 2.3 e.u.\ and 
1.9 e.u.\ used by Stull and Sinke. In any case, the uncertainty in the value 
of $\ln z'$ associated with the choice between $\ln (z-1)$ and $\ln (z-2)$ 
does not exceed 8\%, which is well within the $\sim 20$\% uncertainty of Eqs.\ 
(4),(5). 
 
In ref.\ \cite{prev1}, for three-quarters of the Periodic Table, we 
established that
\beq 
b^2\rho (T_m)=0.64\pm 0.14. 
\eeq 
It then follows from Eqs.\ (3)-(5) with $z'=z-1$ and the numerical values 
of the parameters quoted above, 
that 
\beq
B=
0.0369\pm 0.0093, 
\eeq 
where 
most of the error comes from uncertainty in the value of $\kappa .$ Hence, 
uncertainty in the numerical value of $B$ is about 25\%, 
which is sufficiently small that $B$ can be considered an elemental constant. 

As follows from Eqs.\ (1), (4) with $z'=z-1,$ and Eq.\ (14), 
\beq
L=
0.0308\pm 0.0113, 
\eeq
where we have used $z=8$ and 12 to fix the error bar. Also, from using this 
value of $L$, the modified Leibfried number defined by Eqs.\ (9) and (11) with 
$z'=z-1$ is 
\beq
L'=
0.0337\pm 0.0119, 
\eeq 
where, again, the error reflects the two values of $z.$ These values of $L$ 
and $L'$ are in agreement with the values obtained in ref.\ \cite{Gsch} 
for almost all the elements of the Periodic Table: $L=0.0305\pm 0.0135,$ 
$L'=0.0334\pm 0.0145.$ Although the values of $B,$ $L,$ and $L'$ do agree 
within errors, uncertainties associated with both $L$ and $L'$ are about 35\% 
which is so large that they cannot be considered elemental constants. 

To conclude, we have defined and evaluated the Bragg, Leibfried and modified 
Leibfried numbers in the theory of dislocation-mediated melting. We have 
explained the approximate numerical coincidence of the Bragg and modified 
Leibfried numbers. We have shown that the Bragg number can be considered 
an elemental constant, in contrast to the Leibfried and modified Leibfried 
numbers. In our analysis, the quantity $\ln z'$ is identified with 
Gschneidner's constant $K.$ The factor $\ln z'$ can only arise naturally 
in a theory of line-like defects.

\end{document}